\documentclass{article}
\usepackage[utf8]{inputenc}
\usepackage[utf8]{inputenc}
\usepackage{a4,latexsym,graphicx,color}
\usepackage{amsmath}
\usepackage{subcaption}

\usepackage[numbers,sort&compress]{natbib}

\title{Effect of Contact Angle on Capillary Pressure and Liquid Recovery from Angular Pore Channels}

\author{Afshin Davarpanah\footnote{Correspondence to: afd6@aber.ac.uk},  Simon Cox, \\
Department of Mathematics, Aberystwyth University,\\ Aberystwyth SY23 3BZ, United Kingdom}

\date{}

\begin{document}

\maketitle

\begin{abstract}
We predict the {volume} of liquid recovered from different-shaped prismatic channels following gas displacement. This recovery factor depends strongly on the contact angle at which the gas-liquid interfaces meet the walls of the channel. {We show that scaling the contact angle by a critical value, dependent on the channel shape, leads to a master curve for the recovery factor.} Our numerical simulations {with the Surface Evolver} are supported by theoretical calculations for the capillary entry pressure in a channel with a {triangular or} rectangular cross-section and for the interface shape in a {``scalloped"} triangular channel formed between three parallel cylinders.
\end{abstract}


\section{Introduction}

The phenomenon of immiscible displacement is a crucial factor in enhanced oil recovery from porous media. In pore-scale networks, fluid interfaces generate capillary forces that can trap hydrocarbons and hence reduce the recovery factor~\cite{dullien,castro}. The significant factors in determining capillary behaviour in porous media are the pore structure itself and the wettability. 

Wettability is defined as a fluid's relative tendency to adhere to the solid phase. It is measured by the contact angle at which a fluid interface meets a solid surface at mechanical equilibrium. For contact angles below $\frac{\pi}{2}$, the surface is known as wetting, while for contact angle greater than $\frac{\pi}{2}$, it is called non-wetting.

The contact angle is directly related to the balance between the surface tension $\gamma$ of the {liquid-gas} interface and the difference in surface tension, which we denote $\gamma_w$, between the wetting film on the solid walls and the {liquid}-solid interface inside the liquid. If we denote the contact angle by $\theta$, then 
                 \begin{equation}
                     \cos\theta = \frac{\gamma_w}{\gamma}
                     \label{eq:define_angle}
                 \end{equation}
%This relationship is shown in Figure~\ref{fig:define_angle}. 
When the wall tension $\gamma_w$ is equal to $\gamma$ the contact angle is zero, often referred to as the case of perfect wetting. As $\gamma_w$ decreases the contact angle increases.
We assume here that the contact angle is fixed, and does not depend on the direction of motion of the contact line (i.e. there is no hysteresis between imbibition and drainage). Some authors~\cite{radke92,kagan} include a disjoining pressure in the thin wetting film; we neglect such a contribution here.
%(is this like the original Concus and Finn~\cite{concus} work?)

The usual idealisations of a porous medium consists of packings of spheres~\cite{singh} or cylinders~\cite{sadiq}. To ensure that modeling is capturing processes in real porous materials, it is also important to consider angular pore structures.  In channels with a polygonal cross-section, for example, liquid preferentially collects in the sharp corners, and it is important to determine the shape and size of the corner meniscus as the contact angle varies~\cite{ma,rabbani}. The capillary pressure ($P_c$) is directly proportional to the mean curvature $C$ of the meniscus, as determined by the Laplace-Young equation~\cite{degennes}, $P_c = \gamma C$.

Ma et al.~\cite{ma} investigated the relationship between liquid saturation, corner angle, and contact angle on the meniscus curvature within straight channels with regular polygonal cross-sections. They determined the meniscus curvature for various $n-$sided channels (for example $n=3$ for an equilateral triangle and $n=4$ for a square)  with different contact angles. In terms of the half-corner angle  $\alpha = \pi (n-2)/2n$ (so $\alpha=\frac{\pi}{6}$ for a channel with an equilateral triangular cross-section and $\alpha=\frac{\pi}{4}$ for a channel with a square cross-section), they give the following expression for the capillary pressure when the fluid in each of the $n$ corners occupies an area $A$ of a channel with cross-sectional area $A_T$:
\begin{equation}
    P_c (\theta, \alpha, A) = \frac{\gamma}{R} \sqrt{\frac{\tan\alpha \; A_T\; F(\theta,\alpha)}{n A}},
\label{eq:ma_et_al}
\end{equation}
where $R = \frac{1}{2} W \tan\alpha$ is the radius of the in-circle of the cross-section and the geometric function $F$ is
\begin{equation}
    F(\theta,\alpha) = \frac{ \cos\theta \cos(\theta+\alpha)}{\sin\alpha} 
     -  \left( \frac{\pi}{2}-(\alpha+\theta) \right). 
     \label{eq:Ma-F}
\end{equation}

For example, in a channel with a cross-section in the form of an equilateral triangle (the case $n=3$, $\alpha = \frac{\pi}{6}$ shown in Figure~\ref{fig:shapes}(a)) with side length $W$, $R=\sqrt{3}W/6$ and area $\sqrt{3}W^2/4$, the capillary pressure is 
\begin{equation}
    P_c (\theta, \frac{\pi}{6}, A) = \gamma  
    \sqrt{ \frac{F(\theta,\frac{\pi}{6})}{A}}.
    \label{eq:ma_et_al_tri}
\end{equation}
As the liquid area increases the capillary pressure drops and, since $F(\theta,\alpha)$ is a decreasing function of $\theta$ in the interval $0 \le \theta \le \frac{\pi}{3}$, the capillary pressure also drops as the contact angle increases. When $\theta = \frac{\pi}{3}$ $F(\frac{\pi}{3},\frac{\pi}{6})=0$, the capillary pressure is zero and the interface becomes planar.

We consider a straight 3D channel, to represent a pore body, with a fixed cross-section that is initially filled with wetting fluid (``liquid"). A second, non-wetting fluid (``gas"), is gradually introduced at one end of the channel. The gas must overcome the capillary entry pressure {(or threshold pressure)}, denoted $P_c^e$, before the wetting phase is displaced from the other end of the channel to be recovered. The interface between the phases then moves downstream with a roughly hemispherical form, leaving behind thin triangular regions of {liquid} in the corners, as shown in
{Figure~\ref{fig:shapes}(d)-(f) and in the animation (in the case of a channel with rectangular cross-section) in the Supplementary Information.}
%\cite{movie1},.

It is therefore necessary to determine the capillary entry pressure. For a channel with the same regular polygonal cross-section as above, Ma et al.~\cite{ma}  give
\begin{equation}
    P_c^e =  \frac{\gamma}{R} \left(  \cos\theta + \sqrt{\tan\alpha 
     \left( \cos\theta\sin\theta+\frac{\pi}{2}-(\alpha+\theta) \right) } \right).
    \label{eq:pce}
\end{equation}
The radius of the in-circle $R$ depends on $n$, via $\alpha$, as demonstrated above. Taking the example of an equilateral triangle again, we have
\begin{equation}
    P_c^e = 2 \sqrt{3}  \frac{ \gamma }{W}
    \left(  \cos\theta + \sqrt{\frac{1}{\sqrt{3}}
     \left( \cos\theta\sin\theta+\frac{\pi}{3}-\theta \right) } \right).
    \label{eq:pce_tri}
\end{equation}
{Expressions for $P_c$ for channels with various shapes can be found in~\cite{lenormand,lago_araujo_jcis,lago_araujo_2}.}

{We can then determine the area of liquid remaining in a channel after the drying front has moved far downstream by evaluating the shape of the liquid regions when the capillary pressure of the liquid is equal to the capillary entry pressure. In the case of an equilateral triangular cross-section, for any contact angle, we equate eq.~(\ref{eq:pce_tri}) with eq.~(\ref{eq:ma_et_al_tri}) to find $A(\theta)$.}

{We describe the drainage of channels} with non-uniform or non-polygonal cross-sections, such as a rectangle or the gap between three cylinders, with further analysis and three-dimensional (3D) numerical simulations. We first describe the geometry of the channels under consideration in \S~\ref{sec:geometry}, and outline our numerical method. In \S~\ref{sec:pc_theta} we validate our {simulations} by comparing with predictions for the capillary pressure, which requires that we derive an expression for the capillary pressure in the gap between three cylinders. In \S~\ref{sec:merge} we determine the maximum possible liquid volume in each channel before snap-off occurs and the channel rapidly fills with liquid. And then in \S~\ref{sec:rec_fact} we calculate the capillary entry pressure in these non-polygonal channels, which allows us to determine the liquid recovery factor as a function of contact angle.

\section{Channel Geometry and Numerical Method}
\label{sec:geometry}

\begin{figure}%[h]
\centering
 \includegraphics[width=\textwidth]{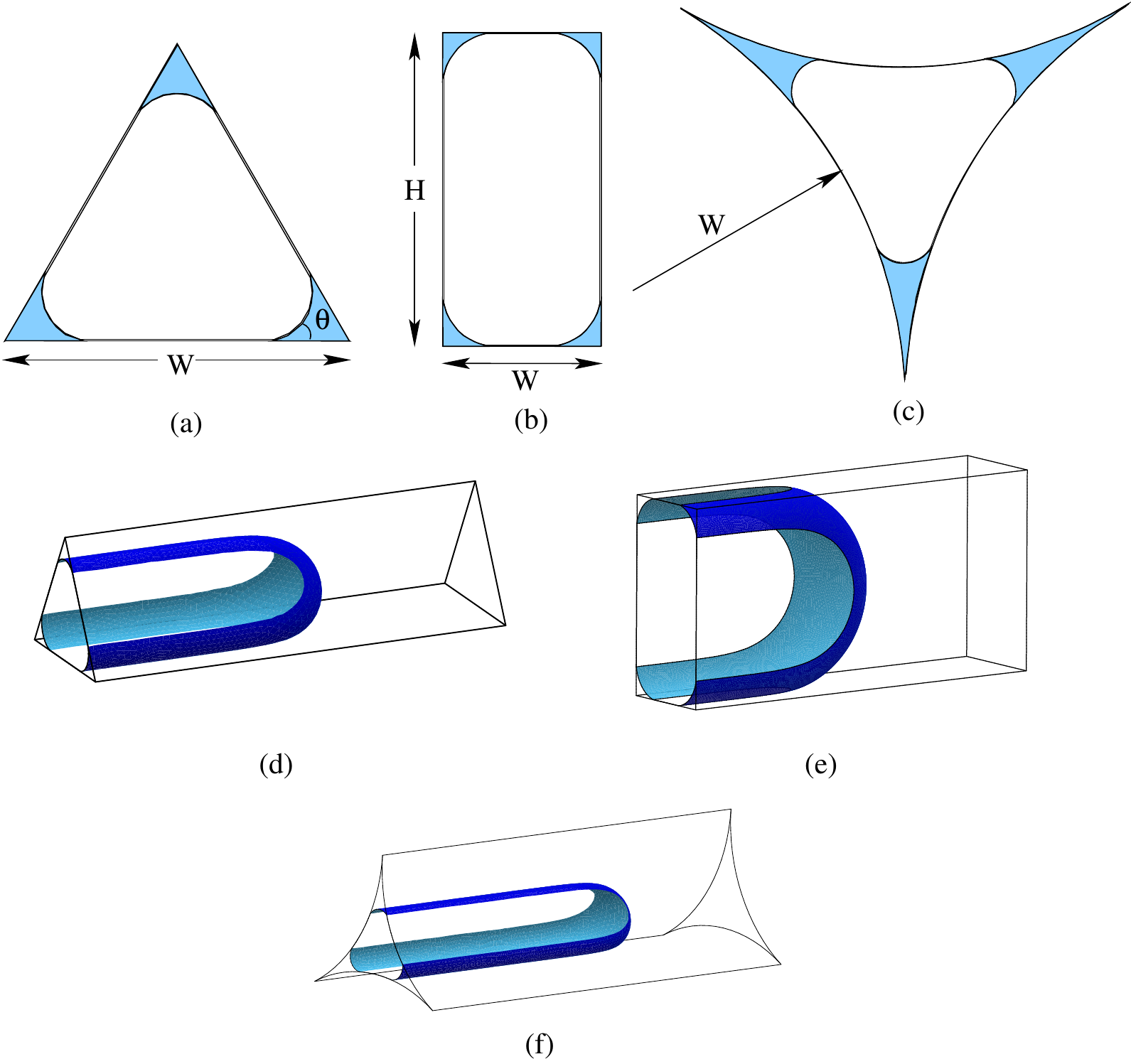}
\caption{Channel shapes considered here, with contact angle $10^\circ$ in each case. (a) Channel with a cross-section in the form of an equilateral triangle, showing the width $W$, contact angle $\theta$ and three regions of liquid in the corners. The height $H=\sqrt{3}W/2$, and the corner half-angle $\alpha=\pi/6$.  (b) Rectangular channel with width $W$ and height $H$. (c) Scalloped triangle between three cylinders of radius $W$. (d) 3D channel corresponding to the cross-section in (a), with length $L$. The interface separating gas from liquid has penetrated  almost half-way along the  channel, moving from left to right, with capillary pressure equal to the capillary entry pressure.  The wetting films on the channel walls behind the moving front are not shown. (e) Equivalent image for a rectangular channel with $H=2W$. (f) Equivalent image for the channel with scalloped triangular cross-section.
%Here  $W=1$, $L=3$ , $\alpha =30^\circ$ and $ \theta?$  is varied through the simulation.
}
\label{fig:shapes}
\end{figure}

We consider channels with a cross-section in one of three shapes shown in Figure~\ref{fig:shapes}:
{
\begin{itemize}
    \item an equilateral triangular with side-length $W$; this shape has no free parameters and allows us to validate our numerical method against known results such as eq.~(\ref{eq:ma_et_al_tri}).
    \item a rectangle with width $W$ and height $H$; we can vary the aspect ratio $H/W$ to determine how the recovery factor depends on this ratio.
    \item the gap between three touching, parallel, circular cylinders~\cite{lago_araujo_2} of radius $W$; we refer to this shape as a scalloped triangle, also considered previously in the context of capillary rise~\cite{princen}, which provides a stringent test for both theory and numerics because of the cusps in the corners of the channel.
\end{itemize}
}

To determine the shape and capillary pressure of the liquid interfaces in the different channels, we perform simulations with the Surface Evolver~\cite{brakke}{, a finite element software designed to predict the static shapes of fluid interfaces. By directly determining interface shape, without the need for a discretisation of space, Evolver offers a fast and efficient alternative to numerical methods such as level sets~\cite{jettesten} and 3D finite element fluid flow calculations with interface tracking~\cite{rabbani}.}

{Surface Evolver provides a surface discretation (into small triangles) and routines for minimizing energy; in this case the energy is the surface area of the interfaces multiplied by their surface tension, and the interface shapes are constrained by the enclosed volumes. The contact angle can be accurately described and quantities associated with the surfaces, such as area and capillary pressure, determined to arbitrary accuracy by repeatedly refining the triangulation into smaller triangles and performing further gradient descent iterations. By slowly incrementing the values of the constraints, such as the enclosed volume, we are able to perform quasi-static simulations of drainage and imbibition, for example.
}

For 2D calculations in the cross-section of a channel there is a single movable interface in each corner of the channel; for a channel size of $W=1$ we refine this interface into short straight segments using three levels of refinement.  In 3D we commence our calculations with the interface roughly half-way along the channel and increase or decrease the gas volume to explore how the capillary pressure varies during steady-state drainage. Without loss of generality we set the interfacial tension to $\gamma=1$ and choose a value of the wall tension $\gamma_w$ to set the desired contact angle via eq.~(\ref{eq:define_angle}). In each case we converge using gradient descent and Hessian steps.

The capillary pressures that we report are in units of $\gamma/W$, i.e. the Laplace pressure. For a dimensional tension $\gamma = 30\times10^{-3} {\rm N/m}$ and pore scale $W= 1 {\rm \mu m}$, we have that $P_c =1$ here corresponds to a capillary pressure of $30\times10^3 {\rm Pa}$. Liquid areas are scaled by the total cross-sectional area of the channel, and liquid volumes by the total channel volume.

\section{Results}

\subsection{Dependence of capillary pressure on contact angle and liquid volume}
\label{sec:pc_theta}

\begin{figure}%[h]
\centering
\begin{subfigure}[b]{0.48\textwidth}
  \includegraphics[width=\textwidth]{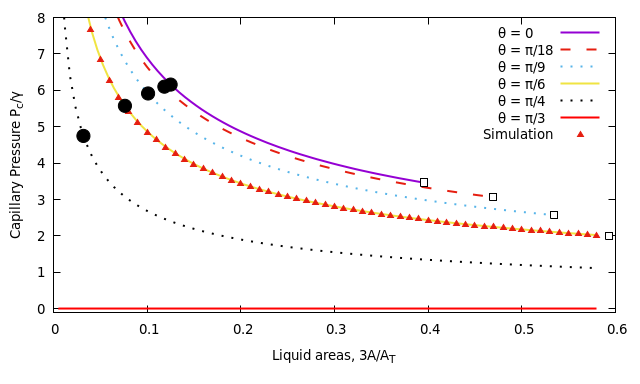}
    \caption{}
\end{subfigure}
\begin{subfigure}[b]{0.48\textwidth}
  \includegraphics[width=\textwidth]{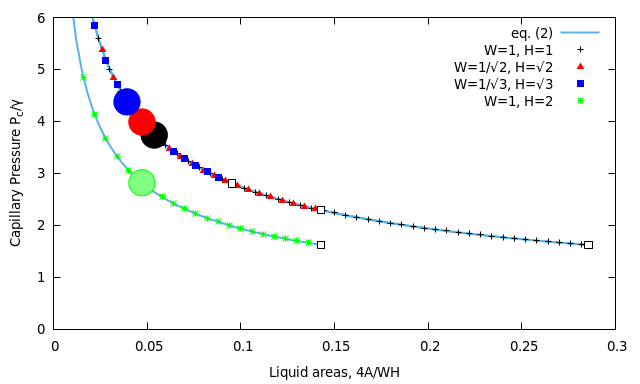}
    \caption{}
\end{subfigure}

\begin{subfigure}[b]{0.48\textwidth}
  \includegraphics[width=\textwidth]{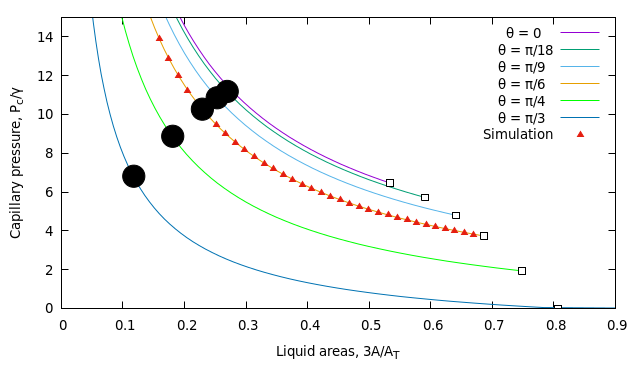}
    \caption{}
\end{subfigure}
\caption{Capillary pressure {\it versus} total liquid area in a 2D cross-section through different channels. Large discs denote the capillary entry pressure; small squares indicate the maximum liquid area at which two (or more) liquid regions merge, from eqs.~(\ref{eq:criticalAandP}) and (\ref{eq:criticalAandP_scall}). (a) An equilateral triangular channel, showing the capillary entry pressure from eq.~(\ref{eq:pce}) superimposed on the lines given by eq.~(\ref{eq:ma_et_al}), for various contact angles. Simulation results are shown as points, validating the simulation technique. The liquid area is scaled by the area of the triangle, $A_T = \sqrt{3}W^2/4$.  (b) Channels with rectangular cross-section, for contact angle $\theta = \frac{\pi}{18}$. Eq.~(\ref{eq:ma_et_al}) with $n=4$ and $\alpha=\frac{\pi}{4}$ gives the lines, the capillary entry pressure is determined by eq.~(\ref{eq:pce_rect}), and the liquid areas are scaled by $A_T=HW$. (c) Channels with scalloped triangle cross-section for various contact angles, showing the capillary pressure from eqs.~(\ref{eq:area_scall}) and (\ref{eq:pressure_scall}) as  lines, compared with the result of numerical simulation in a 2D cross-section in the case $\theta=\pi/3$, and the capillary entry pressure from simulations in a 3D channel. The total channel area is $A_T =  \left(\sqrt{3}-\frac{\pi}{2}\right) W^2$.}
\label{fig:pressure_data}
\end{figure}

Figure~\ref{fig:pressure_data} shows how the capillary pressure varies in the different channels considered here. As the liquid areas increase, the capillary pressures decrease.

Increasing the contact angle also causes the capillary pressure to decrease, leading to a reduction in the amount of liquid left in the channel. At a critical contact angle $\theta_0$, which depends on the shape of the channel, the capillary pressure reaches zero. For a channel with a regular polygonal cross-section the critical angle is $\theta_0 = \frac{\pi}{2}-\alpha = \frac{\pi}{n}$, so for a triangular channel (Figure~\ref{fig:pressure_data}(a)) we have $\theta_0=\frac{\pi}{3}$. A rectangular channel takes the same value as a square channel, $\theta_0=\frac{\pi}{4}$~\cite{rabbani}, while for a scalloped triangular channel, which is not polygonal, it is $\theta_0 = \frac{\pi}{2}$.
%, far from the value for a circular channel of $\theta_0 = 0$.

Compared to the triangular channel, the pressures are lower for given liquid area in a square channel (Figure~\ref{fig:pressure_data}(a) and (b)) because the curvature of the interface required to meet the walls at the required contact angle is less when the angle between the walls is larger. In the scalloped triangular channel (Figure~\ref{fig:pressure_data}(c)) the capillary pressures are therefore even higher.

The capillary pressure of the small regions in the corners of the rectangular channel are identical to those that are found in a square channel up until the point at which they meet (i.e. with capillary pressure given by eq.~(\ref{eq:ma_et_al}) with $n=4$, $\alpha=\pi/4$). For channels with different aspect ratios but the same total area ($HW=1$), Figure~\ref{fig:pressure_data} shows that the data collapse onto one curve, with an increase in the capillary entry pressure as the aspect ratio $H/W$ increases. However, in the rectangular channels with higher aspect ratio the liquid regions meet when their area is smaller. Increasing the channel size (the figure shows the case $HW=2$) reduces the capillary pressures, although the relative liquid area at which the capillary pressure occurs depends only on the channel aspect ratio.

\begin{figure}%[h]
\centering
\begin{subfigure}[b]{0.85\textwidth}
  \includegraphics[width=\textwidth]{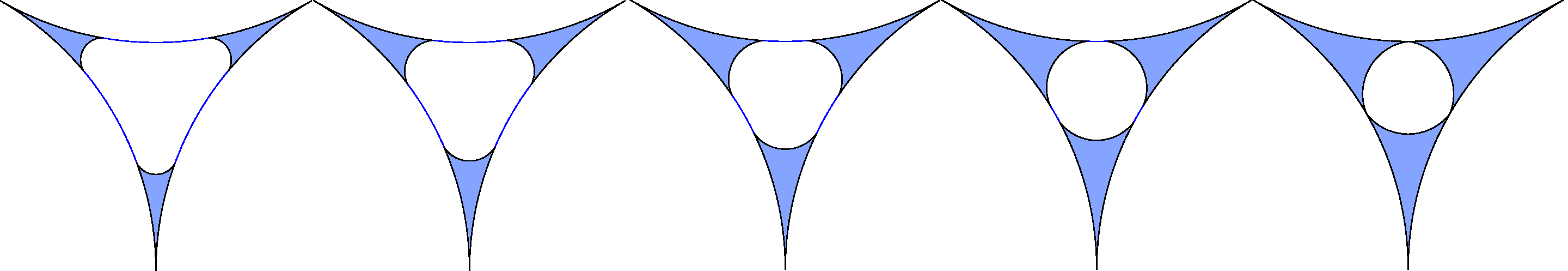}
 \caption{}
\end{subfigure}

\centering
\begin{subfigure}[b]{0.45\textwidth}
  \includegraphics[width=\textwidth]{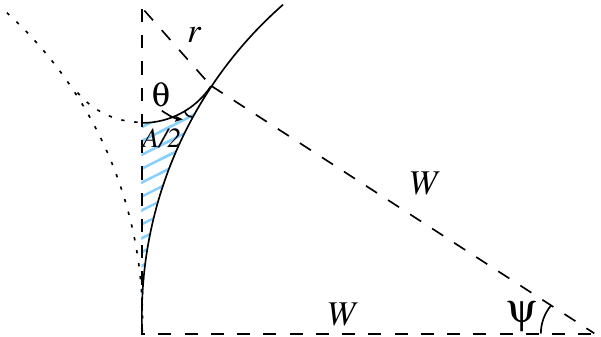}
 \caption{}
\end{subfigure}
\begin{subfigure}[b]{0.45\textwidth}
  \includegraphics[width=\textwidth]{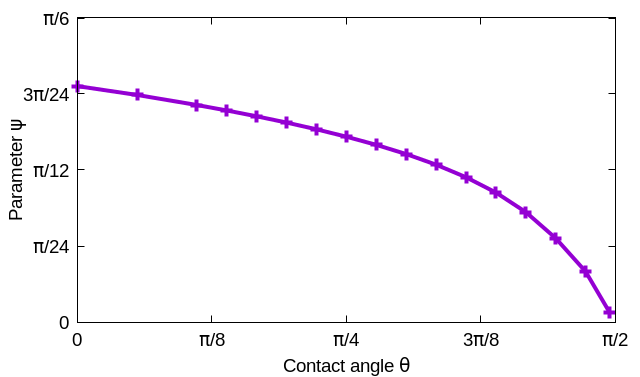}
 \caption{}
\end{subfigure}
\caption{(a) From left to right, increasing liquid areas in the scalloped triangular channel, with $3A/A_T = 0.2,0.3,0.4,0.5,0.58$. The right-most image corresponds to the point at which the liquid regions merge. (b) Geometry of the liquid in one corner of a cross-section through the scalloped triangular channel. The two arcs enclosing the liquid meet at the contact angle $\theta$, and this point on the channel wall subtends an angle $\Psi$ at the centre of the cylinder. The inner interface has radius of curvature $r$, which sets the capillary pressure $P_c$. (c) Solution of eq.~(\ref{eq:psi_scall}) for the value of $\Psi$ at which the capillary entry pressure is reached as the contact angle varies.}
\label{fig:geom_scall}
\end{figure}

For the scalloped triangular channel, we determine the dependence of the liquid area and capillary pressure on the contact angle in terms of a parameter $\Psi$. This is the angle subtended by the point at which the meniscus meets the wall of the channel, measured from the corner of the triangle (see figure~\ref{fig:geom_scall}(b)), and therefore taking values in the range $[0:\pi/6]$. We find
\begin{equation}
    A(\theta, \Psi) = W^2\left[ \tan\Psi - \Psi + 
    \left( \frac{1-\cos\Psi}{\cos(\theta+\Psi)}\right)^2 
    \left( \frac{\sin\theta \cos(\theta+\Psi)}{\cos\Psi}
    +\theta+\Psi - \frac{\pi}{2}\right)
    \right]
\label{eq:area_scall}
\end{equation}
and
\begin{equation}
    P_c(\theta, \Psi) = \frac{\gamma}{W} \frac{\cos(\theta+\Psi)}{1-\cos(\Psi)}.
\label{eq:pressure_scall}
\end{equation}
It is then possible to invert eq.~(\ref{eq:pressure_scall}) to give $\Psi$ in terms of $p_c$, and hence write $A$ in terms of $p_c$. The result is shown in figure~\ref{fig:pressure_data}{, and this appears to agree with the calculations in~\cite{lago_araujo_2} (Table 3).}

\subsection{Merging of liquid regions}
\label{sec:merge}

We now determine the point at which the capillary pressure curves in Figure~\ref{fig:pressure_data} stop, often well before the point at which liquid fills the channel ($n A=A_T$), because two (or more) liquid regions have come into contact, or fused~\cite{ma}. Figure~\ref{fig:geom_scall}(a) shows the point just before this happens, in the case of the scalloped triangle. In other words, we seek the point at which the wetted length on the channel wall is equal to half the length of the wall. 

For a polygonal channel, the wetted length can be written in terms of each liquid area $A$ or the capillary pressure~\cite{ma,helland} as
\begin{equation}
    L = \frac{\cos(\theta+\alpha)}{\sin\alpha} \sqrt{ \frac{A}{F(\theta,\alpha)} } 
    = \frac{\cos(\theta+\alpha)}{\sin\alpha} \frac{\gamma}{P_c},
\end{equation}
where $F$ is defined in eq.~(\ref{eq:Ma-F}) and $A$ is the area of liquid in each corner, as before. When $L$ becomes equal to $W/2$, we find the critical capillary pressure and liquid areas:
\begin{equation}
 P_c^{\rm crit} = \frac{2\gamma}{W}  \frac{\cos(\theta+\alpha)}{\sin\alpha},
 \qquad
 \quad
 A^{\rm crit} = F(\theta,\alpha) \left( \frac{\gamma}{P_c^{\rm crit}} \right)^2 
 = \frac{W^2}{4} \frac{\sin\alpha \; F(\theta,\alpha)}{\cos(\theta+\alpha)}.
 \label{eq:criticalAandP}
\end{equation}

These are marked in Figure~\ref{fig:pressure_data}(a) for the equilateral triangle ($W=1, \alpha=\pi/6)$. For the rectangular channel, merging occurs first on the shortest side. Without loss of generality we can take this to be the width $W$, and then eq.~(\ref{eq:criticalAandP}) applies -- see Figure~\ref{fig:pressure_data}(b).

For the scalloped triangle, eq.~(\ref{eq:criticalAandP}) does not apply. Instead we determine the critical capillary pressure and liquid area from eqs.~(\ref{eq:area_scall}) and (\ref{eq:pressure_scall}) with $\Psi = \pi/6$:
\begin{equation}
 P_c^{\rm crit} = \frac{2\gamma}{W}  \frac{\cos\left(\theta+\frac{\pi}{6}\right)}{(2-\sqrt{3})},
 \qquad
 \quad
 A^{\rm crit} = W^2 \left[
 \frac{1}{\sqrt{3}} - \frac{\pi}{6} + 
 \frac{(7-4\sqrt{3})}{4\cos^2\left(\theta+\frac{\pi}{6}\right)}
 \left( \frac{2}{\sqrt{3}}\sin\theta \cos\left(\theta+\frac{\pi}{6}\right) 
 + \theta - \frac{\pi}{3}\right)
 \right] .
 \label{eq:criticalAandP_scall}
\end{equation}
This is shown in Figure~\ref{fig:pressure_data}(c), again in good agreement with the data.

Following the merging of two liquid regions, the capillary pressure would increase as the total liquid area increased. This is an unstable situation, leading to what is known as snap-off~\cite{rossen} and the channel filling with liquid. We do not consider it further here, but it is important to determine the liquid volumes at which it occurs.

\subsection{Capillary Entry Pressure and Liquid Recovery Factor}
\label{sec:rec_fact}

The data of Figure~\ref{fig:pressure_data} can be summarised by calculating the liquid recovery factor (RF), that is, the relative volume of liquid (wetting phase) removed from the channel:
\begin{equation}
    RF = 1-\frac{nAL}{A_TL} = 1-\frac{nA}{A_T},
\end{equation}
where $A$ is measured at the capillary entry pressure, $P_c^e$, corresponding to the area of liquid remaining in the triangular regions of liquid in the corners of the channel after the {roughly} hemispherical front has passed through the channel (Figure~\ref{fig:pressure_data}).

For the triangular channel we have
\begin{equation}
    RF_{\rm tri} = 2 -  \frac{4\sqrt{3}}{W} \frac{\gamma \cos\theta}{P_c^e},
    \label{eq:RFtri}
\end{equation}
where $P_c^e$ is given by eq.~(\ref{eq:pce_tri}).

For a rectangular channel we have
\begin{equation}
    RF_{\rm rect} = 2 -  \left(\frac{1}{W}+\frac{1}{H}\right)\frac{2\gamma \cos\theta}{P_c^e},
    \label{eq:RFrect}
\end{equation}
{ with the capillary entry pressure  (from ~\cite{lago_araujo_jcis} (eq. 51), re-cast in our notation) given by:}
\begin{equation}
    P_c^e =  \gamma \left( \left(\frac{1}{W}+\frac{1}{H}\right)  \cos\theta + \sqrt{
    \left(\frac{1}{W}+\frac{1}{H}\right)^2 \cos^2\theta - 
    \frac{4}{WH} 
     \left[ \cos^2\theta-\cos\theta\sin\theta-\left(\frac{\pi}{4}-\theta \right) \right] 
     } \right).
    \label{eq:pce_rect}
\end{equation}
This is derived using what Mason and Morrow~\cite{mason} call the ``MS-P" method, after Mayer and Stowe~\cite{mayers} and Princen~\cite{princen}, and improves upon the formula given by Lenormand et al.~\cite{lenormand} for a rectangle with zero contact angle, which omits the term proportional to $4/WH$ (and sets $\cos\theta=1$).

For the scalloped triangle, the expressions required to derive the dependence of the capillary entry pressure on the radius $W$ and contact angle $\theta$ are given by Mayer and Stowe~\cite{mayers} and, in the slightly more general case where the circles need not touch, by Princen~\cite{princen}. These formulae do not admit a closed form solution, but require numerical solution. In our notation, the MS-P method leads to the following nonlinear equation for the critical value of the parameter $\Psi$ (which we denote $\hat{\Psi}$, a function of the contact angle $\theta$) at which the capillary entry pressure occurs:
\begin{equation}
    \left(\frac{1}{\sqrt{3}} - \frac{\pi}{6} + \hat{\Psi} - \tan\hat{\Psi} \right)
    =
    \frac{(1-\cos\hat{\Psi})^2}{ \cos^2(\theta+\hat{\Psi})}
    \left( \frac{\sin\theta \cos(\theta+\hat{\Psi})}{\cos\hat{\Psi}}
    + \frac{\pi}{2}-\theta-\hat{\Psi} \right)
    +\frac{2 (1-\cos\hat{\Psi}) \cos\theta}{\cos(\theta+\hat{\Psi})} \left(\frac{\pi}{6}-\hat{\Psi}\right),
\label{eq:psi_scall}
\end{equation}
and then the capillary entry pressure is given by evaluating eq.~(\ref{eq:pressure_scall}) with $\Psi=\hat{\Psi}$, i.e. $P_c^e  = P_c(\theta,\hat{\Psi})$. We solve eq.~(\ref{eq:psi_scall}) using a numerical root-finding algorithm, the result of which is shown in Figure~\ref{fig:geom_scall}(c), to give the values of $P_c^e$ shown in figure~\ref{fig:pressure_data}(c). The recovery factor is then
\begin{equation}
        RF_{\rm sc-tri} = \left(\frac{6}{\sqrt{3}-\frac{1}{2}\pi}\right) 
        \frac{\gamma}{W P_c^e} \left[\frac{\gamma}{W P_c^e}
            \left( \frac{\pi}{2} - \theta - \Psi \right)
            + \cos\theta \left( \frac{\pi}{6}-\Psi \right)
            \right].
    \label{eq:RFscall}
\end{equation}

\begin{figure}%[h]
\centering
\begin{subfigure}[b]{0.45\textwidth}
  \includegraphics[width=\textwidth]{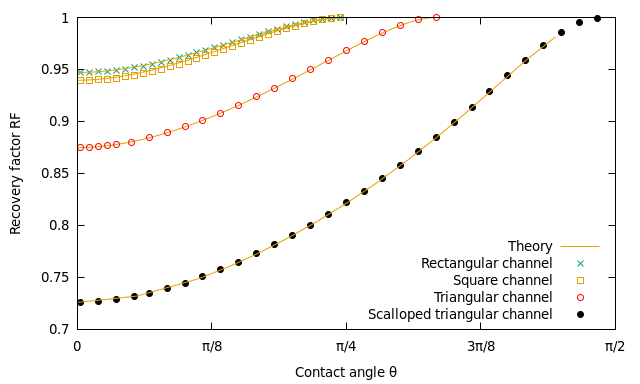}
 \caption{}
\end{subfigure}
\begin{subfigure}[b]{0.45\textwidth}
  \includegraphics[width=\textwidth]{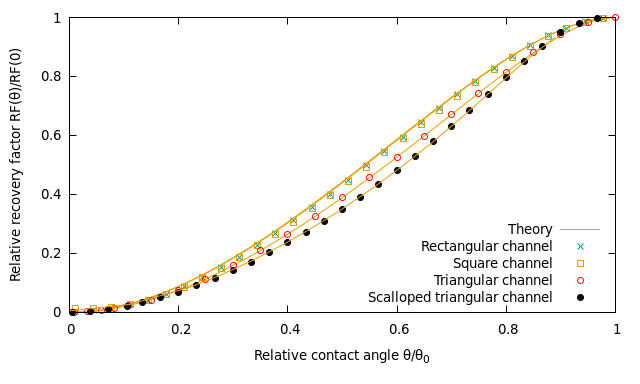}
 \caption{}
\end{subfigure}
\caption{Liquid recovery factor for different contact angles in different shaped channels. The rectangular channel has aspect ratio $H/W=2$. (a) Recovery factor {\it versus} contact angle.  (b) Recovery factor scaled by its value for contact angle $\theta = 0$ {\it versus} relative contact angle, defined as the contact angle relative to the critical contact angle at which 100\% recovery is achieved. Numerical simulations are shown with dots. Lines are from eqs.~(\ref{eq:RFtri}), (\ref{eq:RFrect}) and (\ref{eq:RFscall}).}
\label{fig:recovery}
\end{figure}

Figure~\ref{fig:recovery} shows that the recovery factor depends strongly on the contact angle for all three channel shapes, reaching a value of one (i.e. complete emptying of liquid from the channel) when the critical contact angle $\theta_0$ is reached. Our analytic theory is in excellent agreement with the simulations, although in the case of the scalloped triangle we need a straightforward numerical solution of eq.~(\ref{eq:psi_scall}), or reference to Figure~\ref{fig:geom_scall}(c), to completely determine $RF$ for given contact angle $\theta$.

The volume of liquid remaining in the channel in the case of zero contact angle is greatest for the channel with the scalloped triangular section and smallest for a rectangular channel with large aspect ratio. 
{
Scaling the recovery factor by its value for zero contact angle, as shown in Figure~\ref{fig:recovery}(b), leads to a master curve of the form
\begin{equation}
    RF(\theta) = RF(0) \; \; g(\theta/\theta_0)
\end{equation}
in which the contact angle is scaled by its critical value, which implicitly describes the shape of the cross-section of the channel. We have not determined a precise form for the function $g$, which is roughly sigmoidal.
}

%data by its critical values, indicates that these curves are not equivalent, and that the channel shape is also important. The recovery factor rises more quickly for rectangular channels as the contact angle increases relative to its critical value.

\section{Discussion and Conclusions}

In a cylindrical channel the recovery factor is equal to one, irrespective of the contact angle. The calculations that we have described here for channels with corners give a more realistic idea of what is likely to happen in geological pores. The channel with a cross-section in the form of a scalloped triangle is probably one of the more extreme cases likely to be encountered, with its sharp cusp-like corners, and in the case of perfect wetting (zero contact angle) the recovery factor falls to around 73\%.

Our theoretical predictions have largely {captured the dependence of RF on contact angle for the channel shapes that we consider here. The Surface Evolver simulations  provide useful validation and visualisation of the results, and for more complicated channel shapes are likely to be indispensable.}

The collapse of the data in Figure~\ref{fig:recovery}(b) suggests that for any channel shape for which an estimate of the recovery factor in the case of perfect wetting is available, and if it possible to estimate the critical angle $\theta_0$, then the recovery factor for different contact angles follows a similar curve for all channel shapes, and an approximate prediction can be obtained from this almost universal curve. 

An even greater level of realism will be found by considering channels with a cross-section that varies along its length. This of course brings new theoretical and computational challenges. To make progress in their analysis of flow between packed spheres, Singh et al~\cite{singh} assumed that the interface itself is spherical, while Hilden and Trumble~\cite{hilden03} used Surface Evolver to accurately calculate the liquid shape within a planar array of spheres. {More general situations will surely require more ambitious numerical simulations than those described here, but we expect that the procedure that we have outlined will be required. }

\section*{Acknowledgments}

We thank K. Brakke for the development and support of the Surface Evolver. An AberDOC PhD scholarship from Aberystwyth University is gratefully acknowledged.


\begin{thebibliography}{21}

    %1. 
    \bibitem{dullien}
    Dullien, F.A. Porous media: fluid transport and pore structure. Academic press; 2012. %https://doi.org/10.1016/0300-9467(81)80049-4.
    

    \bibitem{castro}
    de Castro, A.R., Oostrom, M., Shokri, N.. Effects of shear-thinning fluids on residual oil formation in microfluidic pore networks. J. Colloid Interface Sci. 2016; 472:34-43. %https://doi.org/10.1016/j.jcis.2016.03.027 
    
     \bibitem{radke92}
    Wong, H., Morris, S. and Radke, C.J., Three-Dimensional Menisci in Polygonal Capillaries. J. Colloid Interface Sci. 1992; 148:317--336.

    \bibitem{kagan}
    Kagan, M. and Pinczewski, W.V. Menisci in a Diamond-Shaped Capillary. J. Colloid Interface Sci. 2000; 230:452--454.
    %doi: 10.1006/jcis.2000.7077.
    

    \bibitem{singh}
    Singh K., Scholl H., Brinkmann M., Di Michiel M., Scheel M., Herminghaus S., Seemann R. The role of local instabilities in fluid invasion into permeable media. Scientific reports. 2017; 7:444. %https://doi.org/10.1038/s41598-017-00191-y 
    

    \bibitem{sadiq}
    Sadiq T.A., Advani S.G., Parnas R.S. Experimental investigation of transverse flow through aligned cylinders. International Journal of Multiphase Flow. 1995; 21:755-74. %https://doi.org/10.1016/0301-9322(95)00026-T 

    \bibitem{ma}
    Ma, S., Mason, G., Morrow, N.R. Effect of contact angle on drainage and imbibition in regular polygonal tubes. Coll. Surf. A: Physicochem. Eng. Aspects. 1996; 117:273-91. 
    %https://doi.org/10.1016/0927-7757(96)03702-8 
    
    %\bibitem{movie1}
    %See the animation in the Supplementary Information for flow in a channel with rectangular cross-section.
 
   \bibitem{lenormand}
    Lenormand, R., Zarcone, C. and Sarr, A. Mechanisms of the displacement of one fluid by another in a network of capillary ducts. Journal of Fluid Mechanics. 1983; 135:337-53. %https://doi.org/10.1017/S0022112083003110 
    
    \bibitem{lago_araujo_jcis}
   Lago, M. and Araujo, M. Threshold Pressure in Capillaries with Polygonal Cross Section. J. Coll. Interf. Sci., 2001; 243:219--226.
   
    \bibitem{lago_araujo_2}
   Lago, M. and Araujo, M. Threshold capillary pressure in capillaries with curved sides. Physica A, 2003; 319:175--187.
   
    \bibitem{rabbani}
    Rabbani, H.S., Joekar-Niasar, V., Shokri, N. Effects of intermediate wettability on entry capillary pressure in angular pores. J. Colloid Interface Sci. 2016; 473:34--43.
    %https://doi.org/10.1016/j.jcis.2016.03.053
    
   


%    \bibitem{concus}
%    Concus, P. and Finn, R. On capillary free surfaces in the absence of gravity. Acta. Math. 1974; 132:177--198.
    %DOI10.1007/BF02392113

   \bibitem{degennes} 
    De Gennes, P.G., Brochard-Wyart, F. and Qu\'er\'e, D., 2004. Capillarity and wetting phenomena: drops, bubbles, pearls, waves, Springer, New York. 
    
   
    
    
    \bibitem{princen}
    Princen, H.M. Capillary phenomena in assemblies of parallel cylinders: II. Capillary rise in systems with more than two cylinders. J. Colloid Interface Sci., 1969; 30:359--371.
    
    %8. 
    \bibitem{brakke}
    Brakke K.A. The Surface Evolver. Exp. Math. 1992;1:141-65. %https://doi.org/10.1080/10586458.1992.10504253 
  
  \bibitem{jettesten}
  Jettestuen, E., Helland, J. O., and Prodanovi\'c, M., A level set method for simulating capillary-controlled displacements at the pore scale with nonzero contact angles, Water Resour. Res.; 49: 4645-4661.
  %doi:10.1002/wrcr.20334. 
  
    \bibitem{helland}
    Helland, J.O. and Skj{\ae}veland, S.M., Physically-Based Capillary Pressure Correlation For Mixed-Wet Reservoirs From A Bundle-Of-Tubes Model. SPE Journal. 2006; 11:171--180.
 
    \bibitem{rossen}
    Rossen, W.R. A critical review of Roof snap-off as a mechanism of steady-state foam generation in homogeneous porous media. Coll. Surf. A: Physicochem. Eng. Aspects, 2003; 225:1-24.
  
   \bibitem{mason}
    Mason, G. and Morrow, N.R. Meniscus Curvatures in Capillaries of Uniform Cross-section. J. Chem. Soc., Faraday Trans. 1. 1984; 80:2375--2393.
 
  \bibitem{mayers}
   Mayer, R.P. and Stowe, R.A. Mercury porosimetry -- breakthrough pressure for penetration between packed spheres. J. Coll. Sci., 1965; 20:893--911.
 
   
   
    \bibitem{hilden03} 
    Hilden, J.L. and Trumble, K.P. Numerical analysis of capillarity in packed spheres: Planar hexagonal-packed spheres. J. Coll. Interf. Sci., 2003;  267:463--474.
\end{thebibliography}
\end{document}